\documentclass[aps,prd,twocolumn,showpacs,groupedaddress,nofootinbib,longbibliography]{revtex4-1}

\expandafter\let\csname equation*\endcsname\relax 
\expandafter\let\csname endequation*\endcsname\relax 

\usepackage{times}
\usepackage{amsfonts}
\usepackage{graphicx}
\usepackage{amsmath}
\usepackage{amssymb}
\usepackage{color}
\usepackage{subfigure}
\usepackage{natbib}
\usepackage{aas_macros}
\usepackage{ulem}

\usepackage[pdftex,colorlinks,citecolor=black,linkcolor=black,urlcolor=black]{hyperref}

\def\beq{\begin{equation}}
\def\eeq{\end{equation}}
\def\bea{\begin{eqnarray}}
\def\eea{\end{eqnarray}}
\def\ben{\begin{enumerate}}
\def\een{\end{enumerate}}

\def\a{\alpha}
\def\b{\beta}
\def\g{\gamma}
\def\D{\Delta}
\def\e{\epsilon}

\def\g{\gamma}

\def\l{\lambda}
\def\m{\mu}
\def\n{\nu}

\def\O{\Omega}

\def\half{{\textstyle{\frac{1}{2}}}}
\def\fourth{{\textstyle{\frac{1}{4}}}}
\def\w{\wedge}

\def\bz{{\bf z}}

\def\bmE{\boldsymbol{E}}
\def\bmB{\boldsymbol{B}}

\def\bmv{\boldsymbol{v}}
\def\bmx{\boldsymbol{x}}
\def\bmy{\boldsymbol{y}}
\def\bmz{\boldsymbol{z}}
\def\bmJ{\boldsymbol{J}}

\def\cJ{{\cal J}}

\def\cC{{\cal C}}

\def\vphi{\varphi}

\defcitealias{gralla-jacobson2014}{GJ14}

\newcommand{\citeeg}[1]{\citep[e.g.,][]{#1}}

\begin{document}

\title{Nonaxisymmetric Poynting Jets}

\author{Samuel E. Gralla${}^{1,2}$ and Ted Jacobson${}^{1,3}$} 

\affiliation{
$^{1}$Maryland Center for Fundamental Physics \& Joint Space-Science Institute, Department of Physics, University of Maryland, College Park, MD 20742, USA\\
$^{2}$Center for the Fundamental Laws of Nature, Harvard University, Cambridge, MA 02138, USA \\
$^{3}$Perimeter Institute for Theoretical Physics, 31 Caroline Street North, Waterloo, ON N2L 2Y5, Canada} 

\begin{abstract}
The relativistic plasma jets from a misaligned black hole-accretion disk system will not be axially symmetric.  Here we analyze nonaxisymmetric, stationary, translation invariant jets in the force-free approximation where the field energy dominates the particle energy.  We derive a stream equation for these configurations involving the flux function $\psi$ for the transverse magnetic field, the linear velocity $v(\psi)$ of field lines along the jet,  and the longitudinal magnetic field $B_z(\psi)$.  The equations can be completely solved when $|v|=1$, and when $|v|<1$ the problem can be reduced to the pure magnetic case $v=0$ by a ``field line dependent boost''.  We also find a large class of nonaxisymmetric solutions with arbitrary dependence on the longitudinal retarded time. A subclass of these constitutes a novel type of jet that has vanishing electromagnetic pressure $\tfrac{1}{2}(B^2-E^2)$ and requires no external pressure for confinement.  We prove that such self-confinement is impossible when $B^2>E^2$.  Finally, we write down specific solutions approximating numerical results for the nonaxisymmetric jet produced by a spinning black hole in an external, misaligned magnetic field.
\end{abstract}

\maketitle

\section{Introduction}

Plasma jets launched by spinning stars and black holes are ubiquitous 
over multiple scales in the universe, yet their internal structure and the 
mechanisms that launch them remain only poorly understood. 
Part of the challenge arises from observational limitations,
and part from the dynamical complexity of the systems. 
But the universality of the phenomenon suggests the possibility  
that some relatively simple physics is at play, perhaps obscured
by the complex details of individual cases. 

A gross simplification that may nevertheless capture some essential physics  
is the force-free approximation, in which it is assumed that the stress-energy 
tensor for the electromagnetic field is conserved since the
4-momentum exchanged with charges is negligible compared with that 
stored in the field. The field strength in a force-free plasma is derived from a
potential as usual, and satisfies the nonlinear equation expressing 
the vanishing of the 4-force on the 4-current.   
Force-free relativistic jets have been considered in a number of previous studies (e.g. \cite{appl-camenzind1993,tchekhovskoy-mckinney-narayan2008,lovelace-kronberg2013} and references therein).

The simplest setting for jet studies is to assume stationarity, axisymmetry, and translational symmetry along the symmetry axis.  This describes steady jets whose width does not change appreciably over their length, such as seen in pictor A and some other FR II type jets.  Relaxing the assumption of translational symmetry allows the jet to expand appreciably over its length, as seen in FR I type jets such as the famous M87.  With all three symmetries the equations can be completely integrated, and with only stationarity and axisymmetry they can be reduced to a nonlinear stream equation using the Grad-Shafranov approach.  Previous analytic studies of jets have focused on these two cases.

In this paper we will consider a third case, where we relax the assumption of axisymmetry, retaining stationarity and translation invariance.  This describes non-expanding jets whose internal fields are nonaxisymmetric.  Such nonaxisymmetries are natural when the jet launching region is not axisymmetric, such as when a spinning black hole is misaligned relative to its accretion disk.  Some recent observations of Faraday rotation gradients in jets (Fig.~2c of Ref.~\cite{gabuzda-reichstein-oneill2014}) suggest distinct regions of different magnetic structure within the jet, 
as would be the case with nonaxisymmetric fields.

Previous studies of nonaxisymmetric jets have been numerical.  In the simplified case of a spinning black hole in a misaligned external magnetic field, Ref.~\cite{palenzuela-etal2010} found a rather interesting nonaxisymmetric jet structure, with two opposing helical structures within a single jet region (see their Figs.~5 and 6 and our Fig.~\ref{fig:numjetorthogonal}).  Remarkably, we can semi-quantitatively match the structure of the jet to an exact solution (Sec.~\ref{sec:lightdart} and Fig.~\ref{fig:numjet}), which effectively extends the size of the simulation box to infinity.  The exact solution can serve as a basis for further investigation, for example of stability over large distances or synchrotron emission from particles loaded onto the jet.   In general, we hope that this kind of matching will allow efficient and detailed study of the structures that emerge in simulations, bridging the gap between numerical and analytical work, and ultimately connecting to observation.

Our main tool for studying nonaxisymmetric jets will be a Grad-Shafranov formalism based on the time and space translation symmetries.  We will derive an equation (cf.\ \eqref{stream}) for the stream function of the magnetic field transverse to the jet, the analog of the pulsar equation for the poloidal field lines of a stationary, axsiymmetric magnetosphere.  Appearing in this equation are two invariants along the transverse magnetic field lines, the longitudinal magnetic field $B_z$ and the longitudinal velocity $v$ of field lines (analogous to the toroidal field and field angular velocity, respectively).  The case $v=0$ is the well-studied case of a pure magnetic field, and if $v$ is a constant less than the speed of light $1$, then the field is just a boost of a pure magnetic field.  Surprisingly, we find 
(cf.\ Sec.\ \ref{BFL} and App.\ \ref{sec:boost})
that even when $|v|<1$ is \textit{not} constant, every solution can still be related to a pure magnetic solution by a ``field line dependent boost''.  (A similar construction holds for $|v|>1$, which admits magnetically dominated solutions for sufficiently large $B_z$.)  The pure magnetic solutions (``force-free magnetic fields'') have been extensively studied as magnetic flux rope models in solar and planetary magnetosheres (e.g., \cite{wiegelmann-sakurai2012}).  This analysis shows how to promote any such rope to an infinite family of energy-carrying jets, in general nonaxisymmetric.

When $v=1$ (cf.\ Sec.\ \ref{darts}) the boost cannot be carried out and the solutions can have qualitatively different properties, such as the ability to self-confine (see next paragraph).  In this case the equations are straightforwardly solved, with $B_z={\rm const}$ and \textit{any} stream function $\psi$ providing an exact solution.  Furthermore, the solutions may be generalized to include 
in the stream function arbitrary dependence on the retarded time $t-z$
as well, which is the cylindrical analog of the general class of null-current solutions found in \cite{brennan-gralla-jacobson2013} and \cite{brennan-gralla2014}.  A subset of these solutions was actually found a century ago by Bateman \cite{bateman1913,bateman1923}, who was attempting to develop a theory of charge-carrying radiation, and unaware of any plasma application.  We follow Bateman in calling these solutions ``light darts''.

The intuition that magnetic fields possess positive pressure suggests that a force-free jet cannot ``self-confine'' and instead must be held in by an external medium, such as a gas or an ambient magnetic field.  The electromagnetic virial theorem precludes stationary, compact three-dimensional blobs of force-free plasma (cf.\ App.\ \ref{sec:virial}), but the case of confinement in only two directions, relevant to jets, is more subtle.  We find 
(cf.\ Sec.\ \ref{Confinement})
that force-free jets \textit{can} self-confine if the field strength is allowed to become null, $F^2\propto (B^2-E^2)=0$.  This can be understood in terms of pressure balance if one regards $|F^2|/4$ as the electromagnetic pressure, 
which vanishes in the null case.  We give a simple solution representing a self-confined jet of this nature. 
We prove that when magnetic domination is assumed, there are no jets (with our symmetries) compactly supported in transverse directions.

We make use of the formalism introduced by Uchida \cite{uchida1997general,uchida1997symmetry}, and further 
developed in \cite{gralla-jacobson2014}, in which the field strength tensor is described by a 2-form $F=d\phi_1\w d\phi_2$, the wedge product of differentials of a pair of scalar fields.  This methodology is extraordinarily efficient.  
With the exception of the discussion of the virial theorem in the appendix, we restrict to flat spacetime.  We use Minkowski coordinates $t,x,y,z$, with signature $({-}{+}{+}{+})$ and orientation defined by the 
volume element $\e=dt\w dx\w dy\w dz$.
Greek indices run over all coordinates, lower-case Latin indices run over 
spatial coordinates $x,y,z$, and upper-case Latin indices run over the transverse coordinates $x,y$.
For the transverse space we also use polar coordinates $r,\varphi$.
Partial derivatives are sometimes denoted with a subscript comma, 
e.g. $\psi_{,r} \equiv \partial\psi/\partial r$.
The ``square" of a tensor is formed using the spacetime metric, e.g.
$F^2=g^{\a\mu}g^{\b\nu}F_{\a\b}F_{\mu\nu}$. The notation $U\cdot F$ 
denotes
the contraction of a vector $U$ with the first index of the form $F$, in tensor
component notation $U^\a F_{\a\b}$. 
We use Heaviside-Lorentz units, and set $c=1$.

\section{Degenerate, stationary, translation invariant fields}

Force-free fields are defined by the condition that the Lorentz force density 
$j\cdot F=0$ vanishes, where $j$ is the current 4-vector and 
$F$ is the electromagnetic field strength 2-form.  If $j\ne0$ 
this implies that $F$ is degenerate, i.e.\ it is the wedge product of two 1-forms.
(The same property holds in ideal MHD, where the electric field vanishes in the
frame of the plasma 4-velocity $U$, i.e.\ $U\cdot F=0$.  Degeneracy is
equivalent to $\bmE\cdot\bmB=0$.)  
Faraday's law\footnote{The time-space components of $dF=0$ correspond to 
$\partial_t\bmB +\nabla\times \bmE=0$, while the purely spatial component 
corresponds to $\nabla\cdot\bmB = 0$.}   $dF=0$ then implies that 
any degenerate electromagnetic field may be expressed as 
\beq\label{FEP}
F=d\phi_1 \wedge d\phi_2
\eeq
in terms of ``Euler Potentials'' $\phi_1$ and $\phi_2$.
In the magnetically dominated ($F^2>0$) case the surfaces 
on which $\phi_1$ and $\phi_2$ are both constant 
may be thought of as worldsheets of magnetic field lines \cite{carter1979,uchida1997general,gralla-jacobson2014}.  
  We call these surfaces ``field sheets''.  The intersection of a field sheet with a constant-$t$ hypersurface is an ordinary lab frame magnetic field line. A frame in which the electric field vanishes at a point has 4-velocity 
tangent to the field sheet there.  

When symmetries are present, the Euler potentials can be assumed 
without loss of generality to take restricted forms \cite{uchida1997symmetry,gralla-jacobson2014}.  Here we are interested in the case of 
two symmetries, time and space translation, generated by 
commuting vector fields $\partial_t$ and $\partial_z$.  
Faraday's law then implies that 
the longitudinal electric field $F_{tz}$ is constant, and 
we will consider the case where  it is zero. 
We also restrict attention here to configurations
with nonzero transverse magnetic field 
(i.e., $F_{xz}$ and  $F_{yz}$ are not both vanishing).\footnote{Configurations
with zero transverse magnetic field have no $z$-term in $\phi_2$ \eqref{EP}. 
These can be obtained by setting $v={\rm const}$
and taking the limit $v\rightarrow\infty$ with $v\psi$ and $B_z$ held fixed.}
For such fields we may choose the potentials to have the form
\begin{equation}\label{EP}
\phi_1 = \psi(x,y), \quad \phi_2 = \psi_2(x,y) + z - v(\psi) t.
\end{equation}
A magnetic field line with $\phi_1=\psi = {\rm const}$  moves in the $z$-direction 
with velocity $v(\psi)$. We 
refer to $v$ as the velocity of 
the field line.  This is analogous to the more familiar angular velocity of field lines in the stationary
axisymmetric case. 
The field strength \eqref{FEP} associated with \eqref{EP} is
\begin{equation}\label{F}
F = B_z dx \wedge dy + d\psi \wedge ( dz - v\, dt ),
\end{equation}
where the longitudinal magnetic field $B_z$ is related to the potentials by 
$B_z dx \wedge dy = d\psi \wedge d\psi_2$.  

If $v=0$ then \eqref{F} is a purely magnetic field. If 
$v$ is constant and $v^2<1$, then one may eliminate $v$ by boosting into the frame moving with speed $v$ in the $z$-direction, in which case the field is again purely magnetic.  
In fact, even for non-constant $v^2<1$ one may eliminate $v$ by $\psi$-dependent boost, 
as described in Sec.~\ref{BFL} and App.~\ref{sec:boost} below. 
If $v^2>1$ then a similar change of variables can 
be made, with suitable adjustments for the fact that
$1-v^2<0$.  In the special case $v^2=1$ the variable change cannot be carried out, but in some sense $v$ is already eliminated.  This case is of considerable interest to us in this paper.  

\subsection{General properties of translation invariant jets}
Given a curve $\cC$ in the $x$-$y$ plane, the 
magnetic flux through the surface $\cC\times\Delta z$ is $\Delta\psi \Delta z$,
where $\Delta \psi = \int_\cC d\psi$ is the change of $\psi$ along $\cC$.  We may always shift $\psi$ by a constant without affecting the field \eqref{F}, and we will take $\psi$ to vanish at the origin. This makes $\psi$ the ``flux function'' in the sense that $\psi(x,y)$ gives the magnetic flux (per unit length along $z$) across any curve connecting $(x,y)$ to the origin.
The invariant $F^2=2(B^2-E^2)$ is given by
\begin{equation}\label{F2}
\tfrac{1}{2} F^2 = B_z^2 + (1-v^2)(\nabla \psi)^2.
\end{equation}
%
The field is magnetically dominated if $B_z^2 > (v^2-1)(\nabla \psi)^2$,
which always holds if $v^2<1$.
The electric and magnetic field components are given by $B_z$ and
\begin{align}
B_x &= - E_y/v = \psi_{,y} \label{x} \\
B_y &= E_x/v = - \psi_{,x} \label{y} \\
E_z&=0,\label{z}
\end{align}
or in vector language as $\bmB=\nabla \psi \times \hat{\bmz} + B_z \hat{\bmz}$ and $\bmE = - v \nabla \psi = - v \hat{\bmz} \times \bmB$. The ``drift velocity" $\bmv_{\rm D}=\bmE\times\bmB/|B|^2$ 
is
\beq
\bmv_{\rm D} = \frac{v}{B_z^2 + (\nabla\psi)^2}
\Bigl(-B_z\psi_{,y}\,\hat{\bmx} + B_z\psi_{,x}\,\hat{\bmy}+(\nabla\psi)^2\,\hat{\bmz}\Bigr).
\eeq
This is the smallest velocity, measured relative to $\partial_t$,  of a frame in which the electric field vanishes (e.g., \cite{gralla-jacobson2014}). The drift velocity is always less than or equal to $v$, 
\beq
{v_{\rm D}}/{v}={B_T}/{B},
\eeq
where $B_T$ is the magnitude of the transverse field.  

To find the current we first evaluate the dual of \eqref{F}, 
\beq
*F  = B_z dt \wedge dz + \star d\psi \wedge ( dt - v dz ), \label{*F}
\eeq
where $\star$ indicates duality on the transverse ($xy$) subspace.
(The derivation makes use of  Eq.~(A13) of \cite{gralla-jacobson2014}.)
The current 3-form $J=d*F$ is
\beq
\label{J}
J =  dB_z\w dt\w dz + [(\nabla^2\psi)\,dt  - \nabla\cdot(v\nabla\psi)\,dz]\w dx\w dy.
\eeq
In particular, the charge density $j^t$ and current density $j^z$ in the $z$-direction are 
\begin{align}
j^t &= -\nabla\cdot(v\nabla\psi) \label{charge} \\
j^z&=-\nabla^2\psi.\label{zcurrent}
\end{align}
(These are the coefficients in $J$ of $\partial_t\cdot\e=dx\w dy\w dz$ and $\partial_z\cdot\e=-dt\w dx\w dy$.)  The charge density is the divergence of the electric field $\bmE = -v\nabla\psi$.  
The transverse currents are 
\begin{align}
j^x &= \partial_y B_z \label{xcurrent} \\
j^y &= -\partial_x B_z. \label{ycurrent}
\end{align}

The energy, linear $z$-momentum, and angular $z$-momentum current 3-forms can be 
expressed as the Noether current
$\cJ_\xi = -(\xi\cdot F)\w *F + \fourth F^2 \xi\cdot \e$, where $\xi$ is the appropriate
spacetime Killing vector, respectively $\partial_t$, $-\partial_z$, and $-\partial_\vphi = -x\partial_y+y\partial_x$
(cf.\ Appendix E  of \cite{gralla-jacobson2014}).
Using the above expressions we find for the corresponding fluxes per unit time per unit area in the $z$-direction,
\begin{align}
\mbox{energy flux}&= v (\nabla\psi)^2 \label{energy-flux} \\
{P_z}\mbox{-flux} &=  \half(1+v^2)(\nabla\psi)^2 - \half B_z^2 \\
{L_z}\mbox{-flux} &= B_z r \psi_{,r}.
\end{align}
%
These results could of course instead be obtained using the energy-momentum tensor.


\section{Stream Equation and Solutions}
So far we have assumed only that the field $F$ is degenerate, which holds
either in the force-free setting or in ideal MHD. Now we impose the force-free
condition, which may be expressed as \cite{gralla-jacobson2014}
\beq \label{FF}
d\phi_1 \wedge J =0= d\phi_2 \wedge J.
\eeq
For the case under consideration, 
the first force-free condition is $d\psi \wedge J=0$,
which is equivalent to the statement that the 4-force on the current
$j\cdot F$ has vanishing $z$ component, i.e.\ that 
the field linear momentum density in the $z$ direction is 
conserved. If $v(\psi)\ne0$ it is also equivalent to the 
vanishing of the $t$ component, i.e.\ to conservation 
the field energy density [see Sec. 7.3 of \cite{gralla-jacobson2014} for explanation of the analogous statements
in the axisymmetric case].
Since the transverse subspace is only two dimensional, the
first force-free condition is simply $d\psi\w dB_z=0$, which implies 
\begin{equation}
B_z=B_z(\psi).
\end{equation}
That is, the longitudinal field is constant on transverse field lines.

The second force-free condition $d\phi_2 \wedge d*F=0$ 
then yields the stream (Grad-Shafranov) equation for $\psi$,
\begin{equation}\label{stream}
(1-v^2) \nabla^2 \psi - v v' (\nabla \psi)^2 + B_z B_z' = 0,
\end{equation}
where prime denotes a derivative with respect to $\psi$. If $v^2=1$ everywhere then the general solution can be immediately written down, Eq.~\eqref{light-dart} below.  If $v^2\neq 1$ then the equation can be transformed to a simpler equation by a field redefinition, Eq.~\eqref{stream'} below.  If $v^2=1$ only on individual transverse field lines (``critical surfaces''), then the equation should be solved on the separate domains and matched at the surfaces, as explained in Sec.~\ref{sec:critical} below.

\subsection{$v^2=1\!:$ light darts}
\label{darts}

If $v^2=1$ everywhere, then 
Eq.~\eqref{stream} reduces to the demand that $B_z=\textrm{const}$.  
This yields the solution
\begin{equation}\label{light-dart}
F = B_z dx \wedge dy + d\psi \wedge (dz \pm dt),
\end{equation}
which is force-free for \textit{any} choice of $\psi(x,y)$ and constant longitudinal field $B_z$.  
In fact, the field remains force-free if $\psi$ is allowed to depend on $t \pm z$ in addition, i.e., $\psi=\psi(t \pm z,x,y)$.\footnote{Eq.~\eqref{J} for the current still holds for the time-dependent light dart, for which it reduces to $j^\m=-\nabla^2\psi(1,0,0,1)$.  This manifestly has zero contraction with the field strength \eqref{light-dart}.}
Thus there can be arbitrary time dependence at a given value of $z$, which propagates at the speed of light in 
the $z$ direction.
This is the cylindrical analog of the generalized Michel \cite{michel1973} solutions found in \cite{brennan-gralla-jacobson2013}.
It has a null four-current along the $z$ direction.
The solution \eqref{light-dart} with $B_z=0$ and $\psi=\Omega(x,y)f(z-t)$ 
was previously obtained by Bateman in 1923 \cite{bateman1923} and termed a ``light dart''; it is only fitting that we retain this name for the more general expression \eqref{light-dart}.\footnote{Bateman was interested in formulating a theory of charge-carrying radiation and used these null-current solutions as examples.  While for Bateman the null current represented charges moving at the speed of light, in the plasma context it corresponds to net charge and three-current densities of equal magnitude.}

\subsection{$v^2<1\!:$ Boosted field lines}
\label{BFL}

Solutions with $v^2<1$ everywhere\footnote{An analogous treatment is straightforward for $v^2>1$
by allowing $\gamma$ and $\Phi$ to be imaginary. For simplicity we focus on the case $v^2<1$.} are related to $v=0$ solutions by a field-line-dependent boost.  At the level of the stream equation this constitutes introducing a new potential $\Phi$ by
\beq\label{Phi}
d\Phi = d\psi/\gamma, \quad \gamma \equiv 1/\sqrt{1-v^2}.
\eeq
The stream equation now takes the form
\begin{equation}\label{stream'}
\nabla^2 \Phi + B_z B_z' = 0,
\end{equation}
where now the prime denotes derivative of $B_z(\Phi)$ with respect to $\Phi$.
This equation is identical in form to the stream equation in the purely magnetic (or ``nonrelativistic'') case $v=0$; it is satisfied by the force-free magnetic fields studied in the context of solar physics (e.g., \cite{wiegelmann-sakurai2012}).  Thus we may take \textit{any} force-free magnetic field with our symmetries and promote it to an infinite family of relativistic solutions parametrized by a free function $v(\Phi)$. 
More specifically, any solution of \eqref{stream'} together with any choice $|v(\Phi)|<1$ gives rise to a force-free field
\begin{equation}
F =B_z dx \wedge dy + \g d\Phi \wedge (dz - v\,dt).\label{F'1}
\end{equation}
Conversely, every stationary, $z$-translation invariant force-free solution 
with $|v(\Phi)|<1$ arises in this way from a purely magnetic solution by a field-line-dependent boost. 
A similar observation was made in Ref.~\cite{gourgouliatos-etal2012} for the cylindrically symmetric case, including plasma pressure.
In order to gain some insight into why a field line dependent boost is possible, in Appendix \ref{sec:boost} we 
compute the current of the boosted field, finding that it differs from the boosted current
by something whose contraction with the boosted field vanishes.
\footnote{Perhaps the simplest version of this construction begins with a uniform vacuum magnetic field in the $x$ 
direction, and boosts it in the $z$ direction with a $y$-dependent velocity $v(y)$. This results
in a force-free solution with field $F= \g B_x dy\w(dz-v\,dt)$ and non-vanishing current 
3-form $J= -v'\g^3B_x \,dx\w dy\w(dz-v\,dt)$. }

\subsection{Examples of magnetic solutions}

In this subsection we discuss several pure magnetic, $z$-translation invariant 
solutions, to which an arbitrary 
field line dependent boost can be applied to obtain a solution with Poynting flux.

Though simpler than the general case, Eq.~\eqref{stream'} is still in general nonlinear and 
therefore difficult to solve analytically. 
A nonlinear example is the 
cylindrical flux rope of Gold and Hoyle \cite{gold-hoyle1960}, which
follows from the assumption that the number of $\vphi$ 
turns made by the magnetic field lines
per unit length in the $z$-direction is equal to a constant $\nu$.
Then $\psi_{,r}=-B_\vphi = -2\pi\nu r B_z$, which yields 
the solution $B_z\propto 1/[1+(2\pi\n r)^2]$. A class of nonaxisymmetric solutions (including Gold-Hoyle as a special case) follows from choosing $B_z = e^\psi$, giving rise to the Liouville equation, whose analytic solution is known \cite{tassi-pegoraro-cicogna2008}.  Another nonaxisymmetric, nonlinear example is given in Ref.~\cite{khater-etal2006}.

\subsubsection{Linear stream equation}
\label{linear}

Noting that $B_z B_z' = \half (B_z^2)'$, 
any choice of the form 
\beq\label{Bzlin}
B_z^2 = a + b\Phi + c\Phi^2
\eeq
will reduce \eqref{stream'} to a linear equation
for $\Phi$.  
Since $\Phi$  enters \eqref{F'1} only via 
$d\Phi$ and $B_z$, we may without changing $F$ always shift $\Phi$ by a constant, 
making suitable changes of $a$ and $b$.

If $c$ is non-zero we may eliminate $b$ by a shift of $\Phi$, in which case we have
\begin{equation}
B_z=\pm\sqrt{B_0^2 + k^2\Phi^2},
\end{equation}
for constants $B_0$ and $k$, 
and the stream equation becomes the Helmholtz equation,
\begin{equation}
\nabla^2 \Phi = -k^2 \Phi.
\end{equation}
The solutions regular at the origin are Bessel functions,
\begin{equation}\label{besselsoln}
\Phi_m = J_m(kr) e^{i m \varphi},
\end{equation}
where $r,\varphi$ are cylindrical coordinates and $J_m(x)$ is the Bessel function of the first kind.  The general regular solution is the real part of a sum $\sum C_m \Phi_m$ over such modes, where $C_m$ are complex constants. The transverse magnetic fields corresponding to $\Phi_0$ and $\Phi_1+\Phi_{-1}$ are those shown 
in Fig.\ \ref{fig:numjet} (a) and (b), respectively, multiplied by a cylindrically symmetric window function.

When $v=0$ these solutions satisfy $\bmJ=B_z' \bmB$.  If the proportionality factor $B_z'$ is a constant,
which occurs when $B_0=0$, these are called \textit{linear force-free magnetic fields}.  
Such fields are of special interest since they have minimal energy at fixed helicity  \cite{taylor1974}. The $m=0$ solution with $B_0=0$ is the 
solution of Lundquist \cite{lundquist1950}.

If $c=0$ and $b\ne0$ in \eqref{Bzlin} then we may eliminate $a$ by a shift of $\Phi$, yielding
$B_z^2\propto\Phi$. The stream equation then becomes the Poisson equation with a constant source.
With suitable boundary conditions this could be sensible, but note that it requires that $\Phi$ maintain one
sign within the domain of interest.

If $c=0$ and $b=0$, then $B_z$ is constant, and the stream equation becomes
the Laplace equation,
\begin{equation}\label{pearl}
\nabla^2 \Phi = 0,
\end{equation}
which can be solved by separation of variables or conformal maps.  The solutions regular at the origin are
\begin{equation}\label{mode}
\Phi_m = r^m e^{i m \varphi}.
\end{equation}
The general regular solution is the real part of a sum $\sum C_m \Phi_m$ over such modes, where $C_m$ are complex constants. As with the other solutions discussed above, these can be boosted to obtain
other solutions. 
For any constant $v<1$, the result is a vacuum solution, since the charge and current densities vanish (see Eqs. \eqref{charge}-\eqref{ycurrent}). For variable $v$, they are true force-free solutions. 

\subsection{Critical surfaces}\label{sec:critical}

When $v^2=1$ on individual fieldlines but not everywhere, solutions to the stream equation may be constructed by matching at those field lines, which we call ``critical surfaces'' following convention for the axisymmetric stream equation.  At such surfaces our stream equation becomes
\begin{equation}\label{bc}
(\nabla \psi)^2 = B_z B_z' / (v v'),
\end{equation}
which can be viewed as a mixed-type boundary condition for the second-order equation.  Equipped with solutions on either side of the critical surface (constructed either analytically or numerically), one may attempt to match the solutions via the boundary condition \eqref{bc}.  In analytic approaches, the condition would fix free constants or functions, while in numerical approaches it would be used to iteratively update guesses for the undetermined functions $v(\psi)$ and $B_z(\psi)$, following the method of Ref.~\cite{contopoulos-kazanas-fendt1999}.  Experience with the axisymmetric stream equation suggests that each critical surface will place one functional restriction on $v(\psi)$ and $B_z(\psi)$, so that two critical surfaces would uniquely determine both.\footnote{The status of the boundary value problem for equations with critical surfaces is not completely clear, but heuristic arguments (reviewed e.g.\ in  Sec.~7.4.2 of Ref.~\cite{gralla-jacobson2014}) borne out by numerical experience support this general picture.}

\section{Confinement} 
\label{Confinement}

We define the boundary of the jet as the place where the current $j^\mu$ becomes zero,
    and consider three possible scenarios for confinement:
\begin{enumerate}
\item[A.] external confinement by gas pressure,   \label{external} 
\item[B.]  magnetic confinement by externally sourced $B_0\hat{\bz}$, \label{magnetic}  
\item[C.] self-confinement (only transverse fields or no fields outside).
\end{enumerate}
In the first case we allow the electromagnetic field to be discontinuous at the boundary. In the latter cases, on the other hand, force-free solutions go over smoothly to vacuum solutions, so that no external medium is needed.

\subsection{External confinement}

In this case we allow for a discontinuity in the field $F$. The jump conditions
implied by Maxwell's equations are naturally formulated in terms of the three-dimensional
spacetime volume across which the jump occurs (App.~A3 of \cite{gralla-jacobson2014}). Let $\chi(x,y)=0$ be a curve $\mathcal{C}$ in the transverse plane that bounds the jet.  
Translating $\mathcal{C}$ along 
$z$ and $t$ yields the three-volume $\mathcal{S}$ bounding the jet in spacetime.  
The pullback of $F$ to $\mathcal{S}$
must be continuous 
in order to avoid magnetic monopoles on the entailed current sheet.  The simplest case would be to make the field vanish outside the jet, 
in which case the pullback of $F$ to the jet side of $\mathcal{S}$ must vanish. 
From the form of $F$ \eqref{F} we see that vanishing pullback implies $d\psi \propto d\chi$, i.e., 
$\psi$ must be constant on the boundary.
This is nothing but the statement that the 
magnetic field should be tangent to the boundary.
If the field is discontinuous then a surface charge and/or current is present on the boundary.  This may be characterized by a 
current 2-form $K$ living on $\mathcal{S}$, and is given by the jump in the pullback of $*F$ (Eq.~A23 of \cite{gralla-jacobson2014}).  

For a given discontinuous solution $F$ we can ask what external forces are needed to realize that solution.  To do so, consider an infinitesimal patch of 
$\mathcal{S}$
and infinitesimally thicken this patch in 
the transverse dimension.
Then, for each Cartesian Minkowski coordinate $\nu$, integrate $\nabla_\mu T^{\mu \nu}=0$ over the thickened patch, where $T^{\mu \nu}$ includes the stress-energy of the fields 
as well as the matter present outside.  Integrating by parts 
and taking the thickness to zero yields
\begin{equation}\label{force-balance}
( T^{\mu \nu} n_\mu )_{\rm in} = ( T^{\mu \nu} n_\mu )_{\rm out},
\end{equation}
where $n_\mu$ is the normal vector to $\mathcal{S}$ and in/out labels the value obtained when 
$\mathcal{S}$
is approached from the inside/outside.  This equation now holds at any point 
of $\mathcal{S}$
as a tensor equation (i.e. without the restriction to Cartesian coordinates).

To determine the consequences of \eqref{force-balance} 
it is convenient to use the expression (Sec.~3.2.2 of \cite{gralla-jacobson2014}) 
\beq\label{TGJ}
T^{\m\n}=\fourth F^2(h_{\perp}^{\m\n} - h^{\m\n})
\eeq
for the stress tensor associated with any degenerate electromagnetic field. 
Here $h_{\m\n}$ is the (rank 2) metric on the field sheets, and
$h^\perp_{\m\n}$ is the (rank 2) metric on the orthogonal subspace.
The normal covector $n_\m$ 
vanishes when contracted with any vector tangent 
to $\mathcal{S}$. Since the fieldlines are tangent to $\mathcal{S}$ and 
move along $\mathcal{S}$, the field sheets lie in $\mathcal{S}$, and so 
$h^{\m\n}n_\m=0$. It thus follows from \eqref{TGJ} that 
\beq
( T^{\mu \nu} n_\mu )_{\rm in} = \fourth F^2 n^\n.
\eeq
 This equation applies
on the jet-side of $\mathcal{S}$,
and by \eqref{force-balance} it must therefore also hold outside.  
It indicates that the outside stress-tensor must have no transverse momentum, 
no anisotropic stresses, and must have transverse pressure everywhere equal to 
$\tfrac{1}{4} F^2$.
A simple way to satisfy these requirements is to consider a jet where $F^2$ is uniform on the boundary and surrounded by a uniform isotropic gas with pressure
\begin{equation}
p_{\rm gas}=\tfrac{1}{4} F^2.  
\end{equation}
The appearance of $\tfrac{1}{4} F^2$ as the (transverse) jet pressure is not an accident. 
When $F^2>0$ there is a frame in which the electric field vanishes, and in that frame 
$\tfrac14F^2=\tfrac{1}{2}B^2$ is the standard notion of magnetic pressure used in 
magnetic plasmas. The boost relating that frame to the frame of 
the $(t,x,y,z)$ coordinates is orthogonal to the jet normal direction $n^\mu$, so it does not
change the value of the normal pressure. We note from \eqref{F2} that the toroidal contribution 
to the pressure decreases as $1/\gamma^2$, for a given rest-frame toroidal field, indicating that
a highly boosted jet is more easily confined.

Parenthetically, when $F^2<0$ the field is electrically dominated,\footnote{In the electrically 
dominated case force-free dynamics is ill-defined, as it does not admit a well-posed initial value problem \cite{komissarov2002,palenzuela-etal2011,pfeiffer-macfadyen2013}.}
 and the field sheet metric 
$h_{\mu\nu}$ is spacelike. Then \eqref{TGJ} shows that the transverse pressure is still positive.
Hence the quantity $\tfrac{1}{4}|F^2|$ is the relativistic generalization of magnetic pressure, 
which we refer to as the \textit{electromagnetic pressure}. It vanishes in the null case
$F^2=0$ ($E^2=B^2$). Then the expression \eqref{TGJ} for the stress tensor does not 
apply, and instead one has $T^{\mu\nu}= B^2 l^\m l^\n=E^2 l^\m l^\n$, where $l^\m$ is the null 
vector satisfying $l^\m F_{\m\n} = 0$ and normalized with respect to the observer 4-velocity $u^\m$ 
by $l^\m u_\m=-1$.  

To recap, we have considered the basic requirements imposed by terminating the jet discontinuously.  If the fields are to vanish outside the jet then the jet boundary must be a transverse magnetic field line (level set of $\psi$).  In this case an outside pressure equal to the local value of $\tfrac{1}{4}F^2$ must push in on every piece of the boundary.  A gas of uniform pressure is most natural, suggesting that the condition of uniform $F^2$ be imposed on the boundary, which corresponds to fixing $(\nabla \psi)^2$ to be a constant.   Prescribing the shape of the boundary in advance, one should seek solutions of the stream equation such that $\psi$ and $(\nabla \psi)^2$ are constant on the boundary.\footnote{In elliptic equations one normally only can specify either the value or the derivative on the boundary.  It may be, then, that these conditions cannot in general be imposed. They clearly can be imposed in the axisymmetric case, and it is possible that jets externally confined by isotropic pressure must be cylindrical.  In the magnetic or self-confined case nonaxisymmetric jets are possible, as we demonstrate explicitly below.}  

\subsection{Magnetic Confinement}\label{sec:magconf}

We now consider the case where the fields are continuous, but a uniform magnetic field $B_0\hat{\bz}$ exists outside the jet.  The boundary conditions for the stream equation are then $\partial \psi=0$ and $B_z=B_0$.  If $B_z$ varies within the jet, then on account of $B_z=B_z(\psi)$ the boundary must be a level set of $\psi$.  But this constitutes imposing both Dirichlet and Neumann conditions, which is likely to overdetermine the problem and may impose axisymmetry.\footnote{In a large class of non-linear elliptic equations it is known that the overdetermined boundary problem imposes rotational symmetry (in this case it would be axisymmetry) of the boundary and solution (see \cite{serrin1971} and many subsequent references).}  
In the axisymmetric case (full cylindrical symmetry) it is straightforward, using Eq.~\eqref{niceeq}, to construct magnetically confined axisymmetric solutions.
In the non-axisymmetric case we can make progress in the special case where $B_z$ is constant throughout the jet.  This case was considered in Sec.~\ref{linear} in the pure magnetic case $v=0$, which can be variably boosted to obtain the 
general solution with $v^2 <1$. In this case, the stream equation was reduced to the Laplace equation, whose solutions
cannot have compact support. In the exceptional case $v^2=1$, 
the stream function may be chosen arbitrarily (and in particular to vanish outside some boundary).  Thus it is trivial to construct nonaxisymmetric, magnetically confined jets  using the light dart solution \eqref{light-dart}.

\subsection{Self-confinement}

We have discussed jets that are surrounded by gas with isotropic pressure or by a uniform magnetic field.  We now explore the possibility of jets that are surrounded by neither.  If one imposes $B_z=v=0$ at the boundary, then the electric field and the longitudinal magnetic field vanish outside the jet, leaving only transverse magnetic field.  This field will fall off with distance from the jet, and can therefore be regarded as sourced by the jet (rather than as some external field).  (We do not allow electric fields outside the jet
since charges would be attracted, neutralizing the jet.)
It is straightforward to construct such solutions in the cylindrical case using Eq.~\eqref{niceeq}.
(One can choose $B_z(r)$ to vanish outside some radius $r_0$, and integrate to find $B_\vphi$.
The current will then also vanish outside $r_0$, and $B_\vphi$ will be regular at the origin for suitably
chosen $B_z$.)
We see no reason why nonaxisymmetric solutions could not be constructed as well. 

A more extreme version of self-confinement would be a jet where \textit{no} fields persist outside.  In fact that can happen for null ($F^2=0$) field configurations. In cylindrical symmetry,
\eqref{niceeq} implies that for null fields we must have 
$B_z=0$, and that $B_\vphi= \pm E_r$ is \textit{arbitrary}. This is a special case of
the light dart solution, with $B_z=0$ and any $\psi$ of compact support.
On the other hand, no magnetically dominated force-free fields of the form \eqref{F} can have compact support, nor can they even have finite energy per unit length, as we now demonstrate.  

In the appendix we show using a version of the virial theorem that if a stationary force-free field is translation invariant in the $z$ direction, and has finite energy per unit length,
 then $B_z$ and $E_z$ must vanish. Then magnetic domination implies that $v^2<1$ everywhere (see Eq.~\eqref{F2}), and Laplace's equation Eq.~\eqref{pearl} applies with $\Phi$ real. 
Since all regular solutions to this equation diverge at infinity, the field $F$ cannot be 
compactly supported,
nor can the field have finite energy per unit length. The slowest growing regular solution to Laplace's equation has 
$\Phi\sim r$, 
so the magnetic field energy density does not fall off with radius, and produces a quadratic divergence in the energy per unit length.

The ``no-go theorem" for magnetically dominated self-confined jets 
was proved here using conservation of the stress-energy tensor and the stream equation, both 
of which follow from the force-free condition on the electromagnetic field. 
This result can also be established by an argument using the force-free condition more directly,
which is similar to the argument used to establish the ``no closed poloidal loop theorem" for 
stationary axisymmetric magnetospheres in Sec.~7.5.1 of \cite{gralla-jacobson2014}. 
We present this alternate proof in Appendix \ref{sec:proof} 
since the technique may be of interest in its own right.

Finally, we add that it may be of interest to consider configurations whose energy per unit
length diverges logarithmically with the outer radius, like the magnetic energy of 
an infinite line current. This divergence would be rendered finite by end effects, so such 
a configuration could potentially be of physical interest. 
As an example,  the cylindrical flux rope of \cite{gold-hoyle1960} has log-divergent energy; 
however, not only the field, but also the current density fails to have compact support: 
$B_\vphi\sim 1/r$, $B_z\sim 1/r^2$, $j_z\sim 1/r^2$ and $j_\vphi\sim 1/r^3$. This field may be boosted to
a jet with the same falloff behavior.

\section{Light dart jets}\label{sec:lightdart}

In this paper we have derived a stream equation for
translation invariant
nonaxisymmetric jets, discussed its solutions, and considered some boundary conditions of physical interest.  An exceptional case that appeared is the choice $v^2=1$, which we call a light dart.  The light dart field \eqref{light-dart} is force-free for 
constant $B_z$ and \textit{any} choice of $\psi=\psi(t \pm z,x,y)$, and in particular for $\psi$ of compact support.  If $B_z$ is taken to vanish then we have a fully self-confined jet, which evades our theorem by being null ($F^2=0$), rather than magnetically dominated.\footnote{The current density in a magnetically dominated plasma approximated by a force-free solution is typically thought of as arising from particles moving primarily along magnetic field lines on small spirals.
This picture does not directly apply to null fields, since there is no local Lorentz frame in which the field is purely magnetic, so it is less clear whether a null force-free solution can be supported by particles at all.  The particles must move self-consistently in the null field in the appropriate approximation. This might be possible with ultra-relativistic charged particles of a single sign moving along the null direction. Alternatively, a null current can arise from charges of opposite signs moving with different velocities, but such a distribution seems less realistic for an astrophysical jet.}

If $B_z$ does not vanish then the light dart jet is magnetically dominated, but requires an external confining agent.  One may either take $B_z$ to extend outside the jet (magnetic confinement) or have $B_z$ go to zero discontinuously (external confinement).  In the latter case the jet is surrounded by an an azimuthal current sheet and requires an external pressure of $\tfrac{1}{4}F^2=\tfrac{1}{2}B_z^2$ to support it.  

One particular setting in which light dart jets do seem to appear is in the ``Wald configuration'' \cite{wald1974} of a black hole immersed in an external magnetic field. 
 Vacuum solutions are known analytically for any inclination angle between spin and magnetic field \cite{bicak-janis1985}, while force-free solutions have been obtained numerically \cite{komissarov-mckinney2007,palenzuela-etal2010,nathanail-contopoulos2014}.  The
 force-free simulations seem to produce a stationary, asymptotically translation invariant jet.  If this is the case, 
 then the asymptotic solution should fall into our class.
 
 Indeed it appears that the jets of \cite{palenzuela-etal2010} are well approximated by light darts. 
More specifically, numerical data provided to us by Luis Lehner shows that $E_y/B_x$ differs from unity only at the tenth of a percent level in the jet, so that $v = 1 + O(10^{-3})$.  Note that $v=1$ follows from the assumptions of constant $B_z$ and stream function $\psi$ of compact support (see discussion in Sec.~ \ref{sec:magconf}).
 These assumptions are natural to a first approximation for a Wald configuration, and help explain why the special case $v=1$ appears numerically. 
 
\begin{figure}
\centering
\subfigure[]{\label{fig:numjetaligned}\includegraphics[width=35mm]{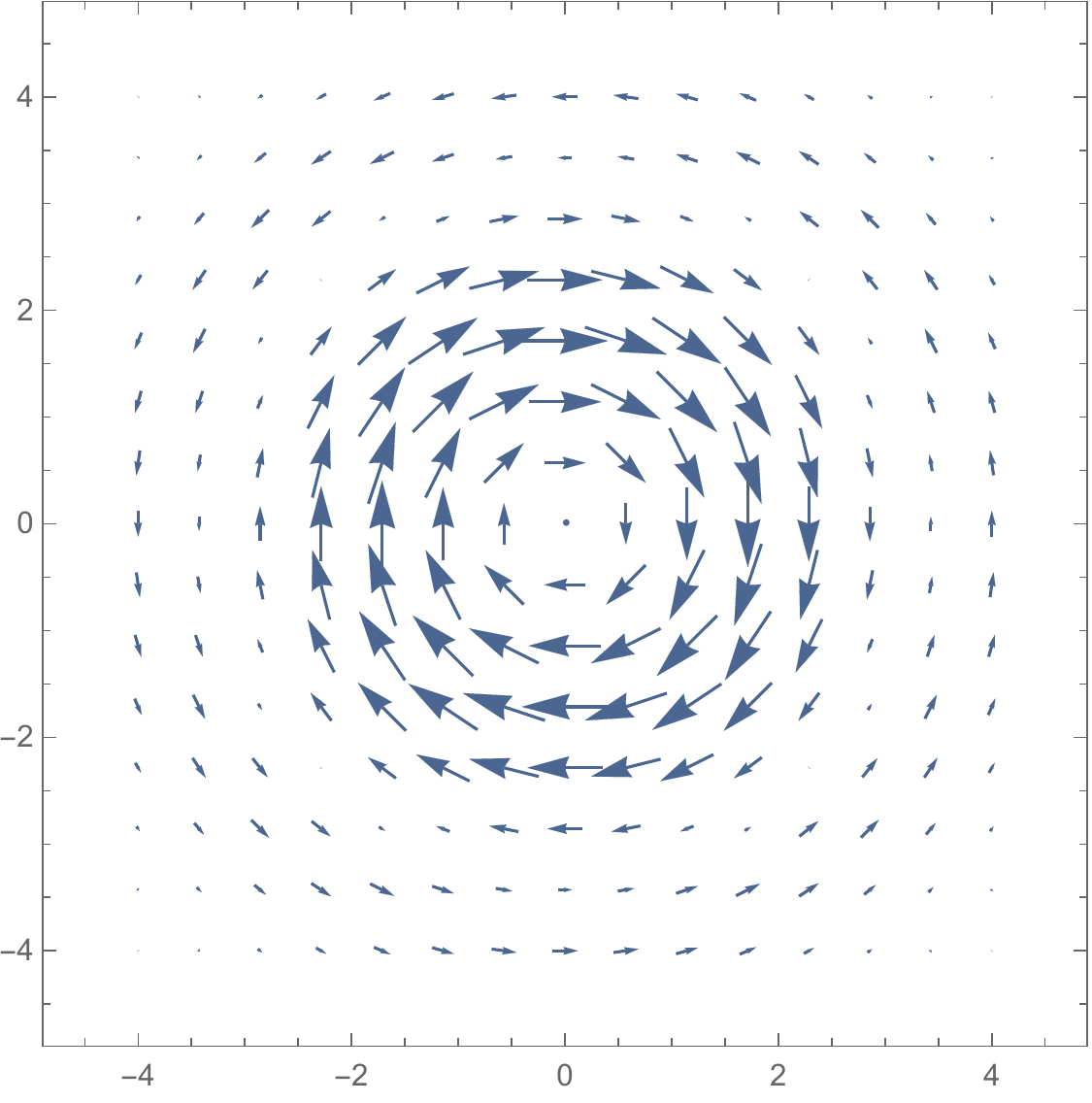}}\ \ \ \ \ 
\subfigure[]{\label{fig:numjetorthogonal}\includegraphics[width=35mm]{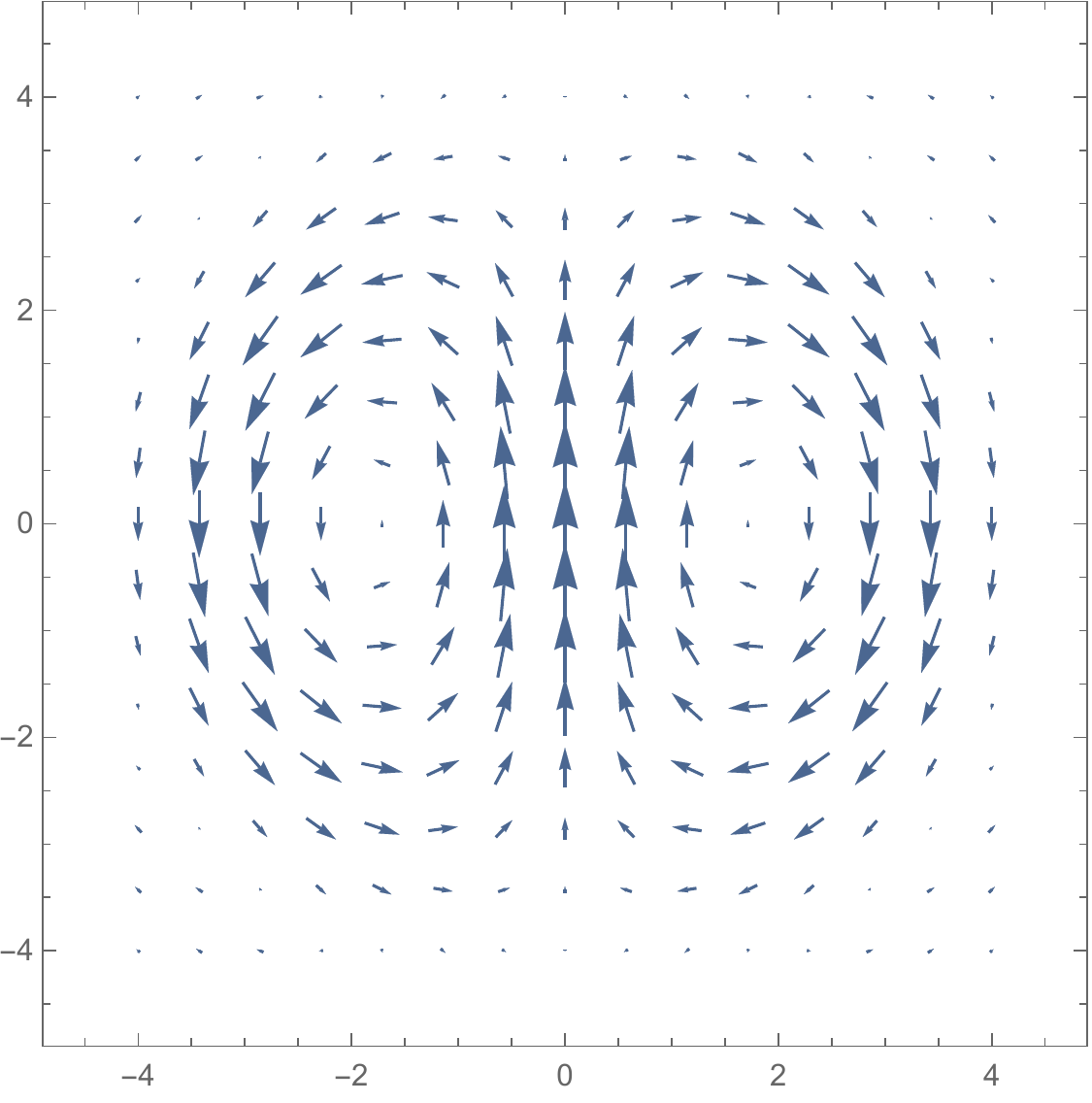}}
\caption{Transverse magnetic field (\ref{x}-\ref{y}) for two light dart jets \eqref{light-dart}, chosen to resemble the jets of a spinning black hole in a magnetic field \protect\cite{palenzuela-etal2010}.  On the left, the choice $\alpha=0$ in Eq.~\eqref{muffin}, which corresponds to to spin-field alignment; on the right, the choice $\alpha=\pi/2$, corresponding to orthogonality.  In each case we have set $\psi_0=-k=-1$, and 
multiplied Eq.~\eqref{muffin} by 
a window function $1-\tanh(r-r_0)$, with $r_0=3.5$ (aligned) and $r_0=4.2$ (orthogonal).}
\label{fig:numjet}
\end{figure}

 To use the light dart solution 
 \eqref{light-dart} to model the jet, it remains to select an appropriate stream function to match the behavior of the transverse fields. The stream function is unconstrained 
 in a light dart, so this matching could be done to arbitrary accuracy at a given value of $z$.
  However, it appears that we can get a rather decent match using the 
 lowest order eigenfunctions of the Laplacian consistent
 with the symmetry of the problem,
\begin{equation}\label{muffin}
\psi = \psi_0 [\cos \alpha\, J_0(kr) + \sin \alpha\, J_1(kr) \sin \varphi],
\end{equation}
where $\alpha$ is the angle between the black hole spin and the asymptotic magnetic field.  
The quantities defining the problem are $B_z$, the black hole mass $M$, the dimensionless black
hole spin parameter $a$, and $\alpha$.  The jet is generated by unipolar induction by the spinning black hole in the magnetic field, so the amplitude $\psi_0$ should scale primarily as the angular velocity of the 
black hole $\O_H(M,a)$ times the flux through the effective conducting region, which includes the 
horizon and perhaps a current sheet generated in the equatorial plane near the black hole.
Since the area of this region scales as $M^2$ (in units with $G=1$),  
this motivates the ansatz
\beq\label{choices}
\psi_0={\tilde\psi}_0 \O_H B_z M^2,\qquad  k=\tilde k/M
\eeq
for the amplitude and the inverse length scale $k$
of the spatial variation within the jet, with dimensionless 
functions ${\tilde\psi}_0(a/M)$ and $\tilde k(a/M)$.
The form \eqref{muffin} describes the jet region only, and should be smoothly cut off outside.  
The aligned ($\alpha=0$) and orthogonal ($\alpha=\pi/2$) cases are plotted in 
Fig.~\ref{fig:numjet}.  These match quite well with Fig.~6 of \cite{palenzuela-etal2010}, and there is similar  qualitative agreement for intermediate alignments (Luis Lehner, private communication).

Using Eq.~\eqref{energy-flux} and some identities for Bessel functions, we find  total power within a radius $r_0=x_0/k$ for the jet \eqref{muffin} to be
\begin{align}
P(\alpha)&=\psi_0^2 \Big[ \tfrac{1}{4} x_0^2 \left(1+\cos^2 \! \alpha \right) \left(J_0(x_0)^2+J_1(x_0)^2\right) \nonumber \\ & \quad -x_0 \cos ^2\! \alpha J_0(x_0) J_1(x_0)-\tfrac{1}{2} \sin ^2\!\alpha \ \! J_1(x_0){}^2 \Big]. \label{power}
\end{align}
The numerical results for the dependence of $P(0)$ on the spin parameter $a$, 
and for $P(\alpha)$ at the spin values $a=0.1,\, 0.7$, are displayed in Fig.~4 of \cite{palenzuela-etal2010}.
The observed spin dependence of $P(0)$ is fit very closely by that of $\Omega_H^2$. 
If we assume that both ${\tilde\psi}_0$ and $x_0= k r_0$ 
are independent of $a/M$, then the spin dependence of \eqref{power}  
is purely via $\Omega_H^2$, in agreement with the numerical results.
Good agreement for $P(\alpha)/P(0)$ is achieved if we adopt for $x_0$ the first zero of $J_1(x)$, $x_0\approx3.83$. Then the $\alpha$ dependence is purely 
via the factor $(1+\cos^2\alpha)$, which is in fact the functional form used by \cite{palenzuela-etal2010} for an analytical fit to their numerical data. Thus the simple form \eqref{muffin}, together with a smooth cutoff at some jet boundary, seems capable of reproducing all of the main features of the numerical results.


Light dart jets do not appear to be restricted to Wald-type configurations.  For example, the externally-confined jets studied in \cite{narayan-li-tchekhovskoy2009}, modeling the jets that arose in numerical simulations \cite{tchekhovskoy-mckinney-narayan2008}, are in the light dart class. This follows from Eq.~(25) of \cite{narayan-li-tchekhovskoy2009}, 
$E_r = B_{\rm \vphi}$, which shows that $v=1$ in their model. 

In these examples the light dart jet has non-zero longitudinal magnetic field $B_z$.  We conclude the paper with 
a speculation
about the 
case $B_z=0$, which has a null field and  can  self-confine.  
Many AGN jets extend for enormous distances (more than one hundred kiloparsecs) 
with degree scale opening angles, and it is a challenge 
to understand this astonishing level of confinement and straightness.
Proposed explanations involve the pressure of the ambient intergalactic medium, a surrounding backflow of shocked gas, transverse expansion of the jet, and shear-induced stabilization  
(e.g., \cite{Hardee2011,Perucho2012,Porth-Komissarov2014}).
To this list we can add the peculiar possibility of null, or rather nearly-null Poynting jets, which are confined by virtue of a balance between electric tension and magnetic pressure.  
More work is 
required before this could be considered a viable explanation.
In particular, the existence and stability of a supporting charge distribution must be established. 
In the meantime, it is amusing to think that Bateman's light darts, conceived by him as a model of light quanta, may actually describe kiloparsec-scale relativistic plasma jets.

\section*{Acknowledgments}
We thank K. Gourgouliatos, M. Grillakis, J. McKinney, and A. Tchekhovskoy 
for helpful comments, and L. Lehner for extended correspondence and sharing of simulation results. 
 This research was supported in part by Perimeter Institute for Theoretical Physics.  Research at Perimeter Institute is supported by the Government of Canada through Industry Canada and by the Province of Ontario through the Ministry of  Research \& Innovation.  S.G. acknowledges support from NASA through the Einstein Fellowship Program, Grant PF1-120082.  S.G. was supported in part by the National Science Foundation under grant No. PHY-1205550.  T.J. was supported in part by the National Science Foundation under grants No. PHY-0903572, PHY-1407744, and PHY11-25915.

\appendix



\section{Virial theorem}\label{sec:virial}

In this appendix we prove a relativistic virial theorem for force-free plasmas with 
both generic three-dimensional shape and for configurations with translation invariance in one direction.
The three-dimensional result is well-known and has been established for general plasma systems (e.g., \cite{laundau-lifshitzEM,laundau-lifshitzClass,finkelstein-rubinstein1964,mobarry-lovelace1986}).   
Such theorems are easily proved in component langauge in Cartesian coordinates for flat spacetime, but here we work with covariant objects insofar as possible, in order to elucidate what spacetime properties are required for the theorems to hold.

Consider any system in flat spacetime with a conserved energy-momentum tensor $T^{ab}$, so that 
for any vector field $\xi^a$ we have the identity
$\nabla_a (T^{ab} \xi_b) = T^{ab}\nabla_a\xi_b$. If $\xi^a$ generates a spatial dilatation in some Lorentz frame 
(for example $\xi^\mu = (0,x^i)$ in Minkowski
coordinates) then (with a suitable normalization) it satisfies $\nabla_{(a}\xi_{b)}=\g_{ab}$, with $\g_{ab}$  the spatial metric, 
and the identity becomes $\nabla_a (T^{ab} \xi_b) = T^{ab}\g_{ab}$.
That is, the spatial trace of the stress tensor is the source of the dilatation current. 
Integrating over a spacetime region $\cal R$, we thus have  
\beq\label{virial}
\oint_{\partial\cal R} T^{ab} \xi_b \,d\Sigma_a= \int_{\cal R} T^{ab}\g_{ab},
\eeq
where the spacetime volume element is implicit in the integral over $\cal R$.
This is a form of the virial theorem. It
can be used  as follows to show that there are no compactly supported, stationary configurations of force-free plasma.

In the Lorentz frame in which the system is stationary, 
choose a spatial region $V$ extending beyond the support of the stress tensor,
and generate a spacetime region $\cal R$ by translating $V$ through some time interval.
Then the entire boundary integral on the left vanishes, because the initial and final time slice
contributions cancel and the rest of the boundary lies outside the support of $T^{ab}$.  It follows that 
$\int_{\cal R} T^{ab}\g_{ab}=0$. The spatial trace of the electromagnetic stress energy tensor is equal to the 
energy density, $T^{ab}\g_{ab}=\half (E^2+B^2)\ge0$, so 
the only way the integral can vanish is if the fields vanish everywhere. 
This derivation can be upgraded to apply to a non-force-free plasma,  since the additional term involving the spatial trace of 
the particle stress tensor is nonnegative as well. 

For a stationary system that is also translation invariant in the spatial $z$ direction, a similar identity can be established
as follows. Let $\xi^a$ now be a dilatation in the two spatial dimensions transverse to the $z$ direction, satisfying 
$\nabla_{(a}\xi_{b)}=h_{ab}$, with $h_{ab}$  the two-dimensional transverse spatial metric. This yields 
an identity like \eqref{virial} with $h_{ab}$ in place of $\g_{ab}$. Now choose the region $\cal R$ to be a transverse
spatial area $A$, translated in both the $z$ and the time directions, and suppose that $T^{ab}$ has no
support outside $A$. It follows that $\int_{\cal R} T^{ab}h_{ab}=0$. The transverse spatial trace of the electromagnetic stress energy 
tensor is $T^{ab}h_{ab}=\half (E_z^2+B_z^2)\ge0$, so the fields must have vanishing $z$-components. In the text, we combine this information with the force-free conditions and the assumption of magnetic domination to conclude that in fact all the fields must vanish
in that case.

The condition of compact support can be weakened to just require finite total energy. 
Consider a boundary $\partial V$ or 
$\partial A$ at fixed spherical or cylindrical radius $r$. The unit normal $n^a$ to the boundary and 
$\xi^a$ are both radial, and $\xi^a = r n^a$ so the integrand 
of the surface integral in the virial theorem
is $\sim T^{rr} r^D$, where $D=3$ for the spherical radius 
and $D=2$ for the cylindrical radius. Finite energy for $D=3$, and finite energy per unit length for $D=2$,  
require $T^{tt} r^D\rightarrow0$. 
The electromagnetic field stress energy tensor
satisfies $|T^{rr}|< T^{tt}$, so finite energy implies that the surface integral in the virial theorem vanishes as $r\rightarrow\infty$.
Thus no finite energy stationary force free configurations exist, and any translation invariant solution with finite energy
per unit length must have $B_z=E_z=0$.
The cylindrical solution of \cite{gold-hoyle1960} comes close to violating this: it has nonzero $B_z$ yet 
its energy per unit length diverges only logarithmically with the outer radius. 

\section{Alternate proof of no self-confined, magnetically dominated jets}\label{sec:proof}

In this appendix we give an alternate proof of the fact that, for a field of the
form \eqref{F}, the magnetic field cannot vanish outside a compact region in the $x$-$y$ plane.
Every smooth such field strength can be represented using a smooth stream 
function $\psi(x,y)$ (the global existence of which is ensured by the Poincar\'{e} lemma), 
which vanishes outside a compact region.
The force-free condition is equivalent to the statement that the two 
3-forms $d\phi_{1,2}\w *F$ are closed. Their integral over any closed 3-surface therefore vanishes, provided that surface bounds a 4-volume in which they are everywhere regular. 
We apply this to the Euler potentials \eqref{EP}, integrating
$d\phi_2\w *F$ over a closed 3-surface formed by translating two 
closed loops in the $x$-$y$ plane
through intervals $\D z$ and $\D t$,
and closing off at the ends of these intervals by filling in the regions between the 
two loops.  
Under the assumption of $z$ and $t$ translation invariance, the contributions to the integral at opposite ends of the intervals cancel, leaving only the integrals over the 
3-surfaces swept out by the two loops.
Choosing one of the loops 
to lie everywhere outside the support of 
the field, its contribution to the integral vanishes as well.
If the other loop is a streamline of constant $\psi$,
then the integral over the corresponding 
constant $\psi$ surface must therefore vanish by itself.
If $d\psi\ne0$ everywhere on the corresponding streamline, and if
$v$ is any vector field satisfying $v\cdot d\psi=1$, that integral is equal to the integral of 
$v\cdot(d\psi\w d\phi_2\w *F) = \half F^2 v\cdot\e$,  where $\e$ is the spacetime volume form (see Eq.~(A7) of \cite{gralla-jacobson2014}). 
But if the field is magnetically dominated ($F^2>0$) then the integral of $F^2 v\cdot\e$ cannot vanish
(since its pullback to the constant $\psi$ surface is nowhere vanishing),
and we reach a contradiction. 
Thus there can be no closed streamline with $d\psi\ne0$ everywhere. 
If every streamline has a point where $d\psi=0$, then
one can move between streamlines without changing $\psi$, so $d\psi=0$ everywhere and the
field is therefore trivial.

\section{Cylindrical symmetry}\label{Cyl}

To relate our formalism to more common approaches assuming axisymmetry, we now discuss the special case in which the field is axisymmetric in addition to being translation invariant and time 
independent. In particular, the spatial symmetry is cylindrical.  
We first work out the relation between the standard axisymmetric 
description and the translation invariant description of the previous section.
Next we present another standard description of this case and some 
particular well-known solutions. 

When $F \cdot \partial_\varphi \neq 0$ (i.e., when there is nonzero poloidal magnetic field) a stationary, axisymmetric, degenerate, regular Maxwell field may be written (cf.\ Eq.~(64) of \cite{gralla-jacobson2014}) as 
\begin{equation}\label{FGS}
F = \frac{-I}{2 \pi r} dr \wedge dz + d \Psi \wedge ( d\varphi - \Omega_F dt ).
\end{equation}
Here $\Psi$ is the magnetic flux through a loop of constant $r$ and $z$, $I=I(\Psi)$ is the electric current through the loop, and $\Omega_F=\Omega_F(\psi)$ is the 
angular velocity of magnetic field lines (as defined by the field sheets). 
In cylindrical symmetry all quantities are functions of $r$ alone
(if we assume the field is regular on the axis), 
and Eqs.~ \eqref{F} and \eqref{FGS} respectively become 
\begin{align}
F & = B_z(r) dx \wedge dy + \psi'(r) dr \wedge ( dz - v(r) dt ) \\
& = \frac{-I(r)}{2\pi r} dr \wedge dz + \Psi'(r) dr \wedge (d\varphi - \Omega_F(r) dt),
\end{align}
(the prime here denotes derivative with respect to $r$).
The relationship is therefore
\begin{align}
B_z(r) = \frac{\Psi'(r)}{r}&,  \quad \psi'(r) = \frac{-I(r)}{2\pi r}, \\
 \psi'(r) v(r) & = \Psi'(r) \Omega_F(r).
\end{align}
In particular we see that the total current within a radius $r$ is $2 \pi r \psi'(r)$.  Thus if a cylindrical jet is to have no net current it must terminate with $\psi'(r)=0$.  Another useful relationship is
\begin{equation}\label{OF}
\Omega_F(r) = \frac{\psi'(r) v(r)}{r B_z(r)}. 
\end{equation}
A jet launched by a rotating conductor (such as a pulsar or an accretion disk) will have 
$\Omega_F(r)$ determined by the angular velocity of the conductor at the footpoint 
of the corresponding magnetic field line.

In the cylindrical case the derivatives with respect to $\psi$ in the 
stream equation \eqref{stream} can be expressed as 
$(1/\psi_{,r})d/dr$. Thus multiplying the stream equation by $\psi_{,r}$ 
converts it to a first order ordinary differential equation involving the electric and magnetic fields,
\beq\label{niceeq}
\frac{dB_z^2}{dr}+\frac{1}{r^2}\frac{d}{dr}\left[r^2(B_\vphi^2-E_r^2)\right]=0.
\eeq
Solutions to this (standard) equation can be found simply by choosing two of the field 
components and integrating to find the third \citeeg{tchekhovskoy-mckinney-narayan2008, narayan-li-tchekhovskoy2009, lovelace-kronberg2013}.

\section{Boosted field line construction}\label{sec:boost}

In the text we observed that if the substitution $d\psi\rightarrow \g d\Phi$ is made
in the stream equation with $v^2<1$, the equation becomes that for a purely magnetic field with
Euler potential $\Phi$. This reveals that all solutions with $v^2<1$ arise from boosting the 
pure magnetic solution  \eqref{F} in a $\Phi$-dependent fashion, i.e.\ independently boosting
the field lines. In this Appendix we attempt to expose why this works
by doing the calculation step by step beginning with the boosted field, rather than
just via the substitution in the stream equation.

We begin with the observation that, for any smooth map 
$\l$ on spacetime $M$,  
the pullback $\l^*F$ of a degenerate electromagnetic 
field $F$ satisfying the covariant Faraday law $dF=0$ is another such field,
since pullback preserves the wedge product and 
commutes with the exterior derivative. In general,
unless $\l$ is a conformal transformation,
$\l^*F$ is not a force-free field, since pullback does not commute with Hodge dual. That is, the new
current $d * (\l^*F)$ is not equal to the pullback $\l^*(d*F)$
of the old current. The case at hand provides an exception: although the pullback of the 
current is not the current of the pullback, they differ by a term 
whose contraction with the field vanishes.

Consider a stationary, $z$-translation invariant
force-free solution 
\beq\label{puremag}
F=d\Phi\w (d\psi_2 + dz) =B_z\,dx\w dy + d\Phi\w dz,
\eeq
with $\Phi=\Phi(x,y)$ and $\psi_2=\psi_2(x,y)$,
and subject it to a $\Phi$-dependent boost $\l[v(\Phi)]$ in the 
$z$-direction,
\beq
\l(x,y,z,t)=\{x,y,\g(z-vt),\g(t-vz)\}.
\eeq
The action of this boost on the basis 1-forms is 
\begin{align}
\l^* dx&=dx,\nonumber\\ 
\l^*dy &= dy,\nonumber\\
\l^* dz&= \g(dz - v\,dt) + \l^z_{,\Phi}\, d\Phi,\nonumber\\
\l^* dt&= \g(dt - v\,dz)+\l^t_{,\Phi}\, d\Phi,
\end{align}
so the pullback of $F$ is given by 
\beq
\l^*F= d\Phi\w [d\psi_2 + \g(dz - v\,dt)].
\eeq
Note that the $d\Phi$ term in $\l^*dz$ does not contribute,
since $d\Phi\w d\Phi=0$, so the value of the pullback $\l^*F$
at each point is the same as for a Lorentz transformation. Introducing the notation
\beq
L dz= \g(dz - v\,dt)\quad  \mbox{and}\quad L dt= \g(dt - v\,dz),\eeq
we have the exterior derivative relations
\begin{align}
d(L dz)&= v'\g^2  Ldt\w d\Phi,\label{dLdt}\nonumber\\
d(L dt)&= v'\g^2  Ldz\w d\Phi,
\end{align}
where $v'=v_{,\Phi}$, and the duality relations
\beq
\star(L dt)=-Ldz,\qquad \star(L dz)=-Ldt,
\eeq
where $\star$ is the Hodge dual on the $t$-$z$ subspace. [In establishing \eqref{dLdt} we have 
used $(\g v)'=\g^3  v'$.]

The boosted field is 
\beq\label{FL}
F_L\equiv \l^*F=  d\Phi\w(d\psi_2 +Ldz)=B_z dx\w dy + d\Phi\w L dz,
\eeq
the dual field strength is 
\beq
*F_L = B_z dt\w dz + \star d\Phi \w Ldt,
\eeq
and the current 3-form $J_L=d*F_L$  is 
\beq\label{JL}
J_L = dB_z \w dt\w dz + d\star d\Phi \w Ldt + v'\g^2 \star d\Phi \w  Ldz\w d\Phi.
\eeq
The last term is proportional to $\star d\Phi \w  F_L$, so it does not contribute to the 
force-free conditions. (It contains both factors of $F_L$. Put differently, its contribution to
the current four-vector is orthogonal to $F_L$.) 
The derivative $v'$ therefore plays no role, so the force-free conditions are the same
as they would have been for a Lorentz-transformed field, and thus (thanks to Lorentz invariance 
of the Hodge dual) they are the same as 
for the original field \eqref{puremag}.

Explicitly, 
the force-free conditions are that the wedge product of $J_L$ with the factors 
$d\Phi$ and $d\psi_2 + Ldz$ of 
$F_L$ \eqref{FL} vanish. Ignoring the third term of \eqref{JL} since it will not contribute anyway,
the first condition, $J_L\w d\Phi=0$, implies $dB_z\w d\Phi=0$,
i.e. $B_z=B_z(\Phi)$, as in the purely magnetic case.
(The 3-form $d\star d\Phi \w d\Phi$ vanishes since both factors
are constructed using only $dx$ and $dy$.)  
The second condition,
$J_L\w (d\psi_2 + L dz)=0$, implies 
\beq\label{J2}
dB_z \w d\psi_2\w dt\w dz+ d\star d\Phi \w Ldt \w Ldz=0.
\eeq
Since $\det \!L=1$ we have $Ldt \w Ldz = \det \!L\, dt\w dz = dt\w dz$, 
so \eqref{J2} is precisely the same equation as would arise 
for the original, pure magnetic field \eqref{puremag}. The stream equation
is therefore unchanged by the field line boost. 
Since $dB_z\w d\psi_2 = B_z' B_z dx\w dy$ and
$d\star d\Phi = \nabla^2\Phi dx\w dy$, it is 
 \beq
 \nabla^2\Phi + (\half B_z^2)' = 0,
 \eeq
in agreement with \eqref{stream'}.

\bibliography{nonaxijet}



\end{document}